\documentclass[twocolumn,aps,showpacs]{revtex4}
\usepackage{float}
\usepackage{graphicx}
\begin{document}
\title{Nonplanar ion-acoustic subsonic shock waves in dissipative electron-ion-pcd plasmas}
 \author{A A Mamun\footnote{Corresponding author:  mamun\_phys@juniv.edu} and B. E. Sharmin}
\affiliation{Department of Physics \& Wazed Mia Science Research Centre,
Jahangirnagar University, Dhaka-1342, Bangladesh}
\begin{abstract}
The dissipative electron-ion-pcd (positively charged dust) plasma, which is observed in both space and laboratory plasmas, is considered.  The basic features of nonplanar cylindrical and spherical ion-acoustic subsonic shock waves in such a medium are investigated by deriving a modified Burgers equation using the reductive perturbation method.  It is found that the stationary pcd species reduces the phase speed of the ion-acoustic waves, and consequently supports the subsonic shock waves due to the kinematic viscosity (acting as a source of dissipation) of the ion species.  It is observed that the cylindrical and spherical subsonic shock waves  evolve with time very significantly, and that the time evolution of the spherical shock structures is faster than that of the cylindrical ones. The implications of the results of the work to space and  laboratory plasmas are discussed.
\end{abstract}
\date{\today}
\maketitle
The  linear 
\cite{Shukla92-dia,DAngelo93-dia,DAngelo94-dia} and nonlinear 
\cite{Bharuthram92so-dia,Popel95so-dia,Popel01ash-dia,Popel01bsh-dia,Popel01csh-dia,Gao01sh-dia,Mamun02npso-dia,Mamun02sh-dia,Shukla03sol-sh,Sahu03npso-dia,Moslem05-sol-dia,Mamun08aso-dia,Mamun08bso-dia,Mamun09npso-dia,Mamun09npsh-dia,Mamun09multi-ion-dia,Shukla09-rev-art,Mamun09-arb-sh,Mamun10npsh-dia,Mamun11dl-dia,Mamun11-discov,Deeba12-DIA,Mamun15so-dia} 
properties of the ion-acoustic (IA) waves  in dusty plasmas play a vital role in understanding small, finite and large amplitude electrostatic disturbances in both laboratory \cite{Barkan96dia-expt,Merlino98dia-expt,Nakamura99sh-expt,Luo99sh-expt,Nakamura01so-dia-expt,Merlino04-dia-expt,Shukla02} and space \cite{Shukla02,Verheest00,Bliok85,Goertz89,Mendis94,Horanyi96} plasmas.
The  linear  properties of the IA waves in dusty plasmas (containing inertialess electron, inertial ion and stationary negatively charged dust species) have been identified theoretically 
\cite{Shukla92-dia,DAngelo93-dia,DAngelo94-dia} as well as experimentally 
\cite{Barkan96dia-expt,Merlino98dia-expt,Merlino04-dia-expt}.  It has been shown here that the phase speed of the IA waves increases with the rise of the charge number density of the stationary negatively charged dust species. This linear feature of the IA waves  in dusty plasma modifies their nonlinear features, and leads to the formation of the modified IA (MIA) supersonic solitary and shock waves \cite{Bharuthram92so-dia,Popel95so-dia,Popel01ash-dia,Popel01bsh-dia,Popel01csh-dia,Gao01sh-dia,Mamun02npso-dia,Mamun02sh-dia,Shukla03sol-sh,Sahu03npso-dia,Moslem05-sol-dia,Mamun08aso-dia,Mamun08bso-dia,Mamun09npso-dia,Mamun09npsh-dia,Mamun09multi-ion-dia,Shukla09-rev-art,Mamun09-arb-sh,Mamun10npsh-dia,Mamun11dl-dia,Mamun11-discov,Deeba12-DIA,Mamun15so-dia}.  

Recently plasmas with positively charged dust (pcd) species, which are observed  in many space (viz. noctilucent clouds \cite{Markus99-NLCs}, Earth's mesosphere \cite{Havnes96,Gelinas98,Mendis04,Mahmoudian13}, cometary tails \cite{Horanyi96}, Jupiter's surroundings \cite{Tsintikidis96}, Jupiter's magnetosphere \cite{Horanyi93}, Saturn rings \cite{Mamun04}, etc.) and laboratory \cite{Samarian01-lab,Khrapak01-lab,Fortov03-ed-lab} plasma systems, has received a great deal of interest in understanding small, finite and large amplitude electrostatic disturbances in both space and laboratory dusty plasmas \cite{Markus99-NLCs,Havnes96,Gelinas98,Mendis04,Mahmoudian13,Horanyi96,Tsintikidis96,Horanyi93,Mamun04,Samarian01-lab,Khrapak01-lab,Fortov03-ed-lab}.

Three principal dust charging mechanisms \cite{Chow93,Rosenberg95,Rosenberg96,Fortov98}, which  make the dust species positively charged,  are: 
\begin{itemize}
\item{The photo-emission of  electrons from the dust surface induced by the flux of photons \cite{Rosenberg96}.} 
\item{The thermionic emission of electrons from the dust surface by the radiative heating \cite{Rosenberg95}.} 
\item{The secondary emission of electrons from the dust surface by the impact of the high energetic plasma particles like electrons and ions \cite{Chow93}.}
\end{itemize}
The dispersion relation for the modified IA (MIA) waves in an  electron-ion-pcd plasma medium (containing inertialess isothermal electron species, inertial cold ion species,  and stationary  pcd species) is given by
\begin{eqnarray}
\omega=\frac{1}{\sqrt{1+\mu}}\frac{kC_i}{\sqrt{1+\frac{1}{1+\mu}k^2\lambda_D^2}},
\label{MIA-dispersion1}
\end{eqnarray}
where $\omega=2\pi f$ and $k=2\pi/\lambda$ in which $f$ ($\lambda$) is the IA wave frequency (wavelength);  $C_i=(z_ik_BT_e/m_i)^{1/2}$ is  the IA speed in which $k_B$ is the Boltzmann constant,  $T_e$ is the electron temperature, and $m_i$ is the ion mass; $\lambda_D=(k_BT_e/4\pi z_i^2n_{i0}e^2)^{1/2}$ is  the IA wave length-scale in which $n_{i0}$ ($z_i$) is the  number density (charge state) of the ion species at equilibrium, and $e$ is the magnitude of the charge of an electron.  We note that for an electron-ion-pcd plasma medium 
$n_{e0}=z_in_{i0}+z_dn_{d0}$ at equilibrium,  where $n_{e0}$ is the electron number density at equilibrium; $n_d$ ($z_d$) is the number density (charge state) of the positive dust species;
$\mu=z_dn_{d0}/z_in_{i0}$. The  dispersion relation (\ref{MIA-dispersion1}) can be interpreted as follows:
\begin{itemize}
\item{The dispersion relation (\ref{MIA-dispersion1})  for the long-wavelength limit (viz. $\lambda\gg\lambda_D$) is
\begin{eqnarray}
\frac{\omega}{kC_i}=\frac{1}{\sqrt{1+\mu}},
\label{MIA-dispersion2}
\end{eqnarray}
which indicate that the phase speed of the MIA waves decreases with the rise of the value of $\mu$.}

\item{It is obvious that $\mu=0$ corresponds to the electron-ion plasma,  and 
$\mu\rightarrow\infty$ corresponds to the electron-dust plasma 
\cite{Mamun04,Samarian01-lab,Khrapak01-lab,Fortov03-ed-lab}. Thus, $0<\mu<\infty$ is valid for the electron-ion-pcd plasma. However, $\mu\gg 1$ is not appropriate for the most electron-ion-pcd plasma, since  the MIA waves are due to the compression and rarefaction, and vise-versa of the ion species in presence of the  pcd species.}

\item{The relation (\ref{MIA-dispersion1})  for the short-wavelength limit (viz. $\lambda\ll\lambda_D$) is $\omega_{pi}$, which is the upper limit of the MIA or IA waves  since it is independent of $\mu$.} 
\end{itemize}
The aim of our present work is to examine the effects of the stationary pcd species, and the new features of the linear MIA waves on the MIA shock waves (SWs) in the electron-ion-pcd plasma under consideration. 

We now investigate the MIA shock waves in a dissipative electron-ion-pcd plasma system, and describe the macroscopic state of this plasma  system as  
\begin{eqnarray}
&&\frac{\partial n_i}{\partial t}
+\frac{1}{r^\nu} \frac{\partial}{\partial r} (r^\nu n_iu_i) = 0,
\label{MIA-b1}\\
&&\frac{\partial u_i}{\partial t} + u_i \frac{\partial u_i}
{\partial r} =-\frac{\partial \phi} {\partial r}+\eta\frac{\partial^2u_i}{\partial r^2},
\label{MIA-b2}\\
&&\frac{1}{r^\nu}\frac{\partial}{\partial r}\left(r^\nu\frac{\partial
\phi}{\partial r}\right)  =(1+\mu)\exp(\phi)-n_i-\mu,
\label{MIA-b3}
\end{eqnarray}
where $\nu=0$ for one dimensional (1D) planar geometry, and $\nu=1\,(2)$ for
cylindrical (spherical) geometry; the electron species has is assumed to obey the Boltzmann law so that $n_e=\exp(\phi)$; $n_e$ ($n_i$) is the   electron (ion) number density normalized by $n_{e0}$ ($n_{i0}$);  $u_i$ is the ion fluid speed normalized by $C_i$;  $\phi$ is the electrostatic  wave potential normalized by  $k_BT_e/e$;  $r$ and $t$ are normalized by  $\lambda_D$ and  
$\omega_{pi}^{-1}$, respectively;  $\eta$ is the kinematic viscosity coefficient (normalized by 
$\omega_{pi}\lambda_D^2$) for the ion species.  The assumption of  stationary pcd species is valid because of the mass of the pcd species being extremely high in comparison with that of the inertial ion species.   

\begin{figure}[htb] 
\includegraphics[width=0.48\textwidth]{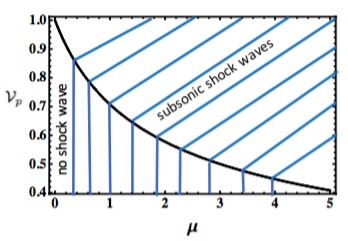}
\caption{The range of the values of $\mu$ and corresponding ${\cal V}_p$  for which the subsonic MIA SWs exist.} 
\label{f1}
\end{figure}

To study the nonplanar SWs in the electron-ion-pcd plasma, we use the reductive perturbation method \cite{Washimi66}, i.e. we stretch the independent variables as \cite{Maxon74,Mamun09,Mamun19}
\begin{eqnarray}
&&\xi=-\epsilon(r+{\cal V}_pt),
\label{str1}\\
&&\tau=\epsilon^2t,
\label{str2}
\end{eqnarray}
expand the dependent variables  as \cite{Washimi66,Maxon74,Mamun09,Mamun19}
\begin{eqnarray}
&&n_i=1+\epsilon n_i^{(1)}+\epsilon^2 n_i^{(2)}+\cdot \cdot \cdot,\\
&&u_i=0+\epsilon u_i^{(1)}+\epsilon^2 u_i^{(2)}+\cdot \cdot \cdot,\\
&&\phi=0+\epsilon \phi^{(1)}+\epsilon^2 \phi^{(2)}+\cdot \cdot \cdot,
\label{np-expan}
\end{eqnarray}
and develop equations in various powers of $\epsilon$,  where $\epsilon$ is the expansion parameter satisfying $0<\epsilon<1$,  and  ${\cal V}_p$ is the phase speed [normalized by $C_i$ 
i.e. ${\cal V}_p=\omega/kC_i$ defined by (\ref{MIA-dispersion2})] of  the MIA waves. 
To the lowest order in $\epsilon$, (\ref{MIA-b1})$-$(\ref{MIA-b3}) give rise to
\begin{eqnarray}
n_i^{(1)}= -\frac{u_i^{(1)}}{{\cal V}_p},
\label{np1}\\
u_i^{(1)}=-\frac{\phi^{(1)}}{{\cal V}_p},
 \label{np2}\\
{\Large {\cal V}_p}=\frac{1}{\sqrt{1+\mu}}.
 \label{np3}
\end{eqnarray}
The variation of ${\cal V}_p$  with $\mu$  is graphically  shown to find the range of the values of 
$\mu$ and corresponding ${\cal V}_p$  for which the subsonic MIA SWs exist. The results are displayed in  figure \ref{f1}. 
It is clear from figure \ref{f1} that the subsonic MIA SWs are formed in the region above the curve (indicated by the horizontal lines), and that no shock wave exists in the region below the curve (indicated by the vertical lines).  
To the next higher order in $\epsilon$, we obtain  
\begin{eqnarray}
&&\hspace*{-10mm}\frac{\partial n_i^{(1)}}{\partial \tau}
-\frac{\partial}{\partial\xi}\left[{\cal V}_pn_i^{(2)}+u_i^{(2)}+n_i^{(1)}u_i^{(1)}\right]-\frac{\nu u_i^{(1)}}{{\cal V}_p\tau}=0,
\label{np4}\\ 
&&\hspace*{-10mm}\frac{\partial u_i^{(1)}}{\partial \tau}-{\cal V}_p\frac{\partial u_i^{(2)}}{\partial \xi}-u_i^{(1)}\frac{\partial u_i^{(1)}}{\partial \xi}
=\frac{\partial \phi^{(2)}}{\partial \xi}+\eta\frac{\partial^2u_l^{(1)}}{\partial \xi^2},
\label{np5}\\ 
&&\hspace*{-10mm}\frac{\partial^2 \phi^{(1)}}{\partial \xi^2}=(1+\mu)\phi^{(2)}-n_i^{(2)} +\frac{1}{2}(1+\mu)[\phi^{(1)}]^2.
\label{np6}
\end{eqnarray}
The use of (\ref{np1})$-$(\ref{np6}) gives rise to a modified Burgers equation in the form
\begin{eqnarray}
\frac{\partial\phi^{(1)}}{\partial
\tau}+\frac{\nu}{2\tau}\phi^{(1)} + {\cal A} \phi^{(1)} \frac{\partial
\phi^{(1)}}{\partial \xi} = {\cal C}\frac{\partial^2 \phi^{(1)}}{\partial
\xi^2}, 
\label{mB}
\end{eqnarray}
where ${\cal A}$  and ${\cal C}$ are nonlinear and dissipation coefficients, and are, respectively,  given by 
\begin{eqnarray}
&&{\cal A}=\frac{3}{2\sqrt{1+\mu}}\left(\mu+\frac{2}{3}\right),
\label{np7}\\ 
&&{\cal C}=\frac{\eta}{2}.
\label{np8}
\end{eqnarray}
\begin{figure}[htb] 
\includegraphics[width=0.48\textwidth]{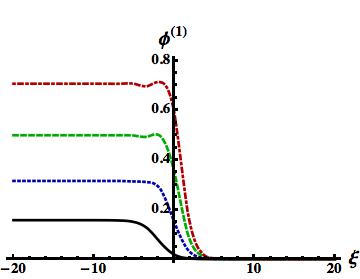} 
\caption{The time evolution of the subsonic MIA  SWs  in cylindrical ($\nu=1$) geometry for ${\cal U}_0=0.1$, $\mu=0.3$, $\tau=-20$ (solid curve),  $\tau=-15$ (dotted curve), $\tau=-10$ (dashed curve), and $\tau=-1$ (dot-dashed curve).}
\label{f2}
\end{figure}
\begin{figure}[htb] 
\includegraphics[width=0.48\textwidth]{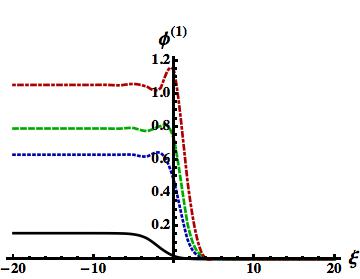} 
\caption{The time evolution of the subsonic MIA SWs  in spherical ($\nu=2$) geometry for ${\cal U}_0=0.1$, $\mu=0.3$, $\tau=-20$ (solid curve),  $\tau=-5$ (dotted curve), $\tau=-2$ (dashed curve), and $\tau=-1$ (dot-dashed curve).}
\label{f3}
\end{figure}
We note that $(\nu/2\tau)\phi^{(1)}$ in (\ref{mB}) is due to the effect of the nonplanar geometry, and that $\nu=0$ or $|\tau|\rightarrow \infty$ corresponds to a 1D
planar geometry. Thus, for a large value of $\tau$,  and for a
frame moving with a speed ${\cal U}_0$,  the stationary shock wave
solution of (\ref{mB}) is \cite{Mamun09} 
\begin{eqnarray}
&&\phi^{(1)}=\phi^{(1)}_0\left[1-\tanh\left(\frac{\xi-{\cal U}_0\tau}{\delta}\right)\right],
\label{solution}
\end{eqnarray}
where $\phi^{(1)}_0$ and $\delta$ are, respectively, the height and thickness of the MID SWs, and are, respectively, given by
\begin{eqnarray}
&&\phi^{(1)}_0=\frac{{\cal U}_0}{{\cal A}},
\label{hgt}\\
&&\delta=\frac{\eta}{{\cal U}_0},
\label{tkn}
\end{eqnarray}
where (\ref{np8}) is substituted into (\ref{tkn}). The direct observations from (\ref{hgt}) and (\ref{tkn}) are as follows:
\begin{itemize}
\item{The electron-ion-pcd plasma system supports the MIA subsonic SWs with
 $\phi^{(1)}>0$, since ${\cal A}>0$ and $\mu>0$ are always valid.}
\item{The height $\phi^{(1)}_0$ of the MIA subsonic SWs depends on ${\cal U}_0$ and $\mu$, but not on $\eta$. It is directly proportional to ${\cal U}_0$ for a fixed value of $\mu$.   On the other hand, it increases with $\mu$ since the phase speed of the MID waves decreases with $\mu$.}
\item{The thickness $\delta$ of the subsonic MIA SWs depends only on $\eta$ and ${\cal U}_0$, and it is directly proportional to $\eta$ for a fixed value of  ${\cal U}_0$, but it is inversely proportional to ${\cal U}_0$ for a fixed value of $\eta$.}
\end{itemize}

We now numerically solve (\ref{mB}) to observe how the subsonic MIA SWs evolve with time from past to present by using (\ref{solution}) as the initial shock pulse. The results for cylindrical ($\nu=1$) and spherical ($\nu=2$) geometries are shown in figures \ref{f2} and \ref{f3}, respectively.  We note that for a large value of $\tau$  (viz. $\tau=20$) we got solid curves [shown in  figures \ref{f2} and \ref{f3}], which represent the subsonic MIA SWs in 1D planar geometry. It is also observe from figures \ref{f2} and \ref{f3} that the height of the cylindrical ($\nu=1$) MIA SWs is larger than that of 1D planar ($\nu=0$) ones,  but smaller than that of spherical ($\nu=2$) ones. Their height (thickness) increases (decreases) as the time pases from past to present.  This is due to the geometry and time dependent term $(\nu/2\tau)\phi^{(1)}$ in the modified Burgers equation.

To summarize, we have considered  the dissipative electron-ion-pcd plasma to show the existence of the subsonic MIA SWs, and to identify their basic features by the numerical analysis of the Burgers equation, which has been derived by employing the reductive perturbation method.  The  limitation of the reductive perturbation method is that it is not valid for arbitrary amplitude SWs.  To overcome this limitation, one has to develop a numerical code to simulate the basic equations (\ref{MIA-b1})$-$(\ref{MIA-b3}). This type of numerical simulation  will be able to show the time evolution of arbitrary amplitude SWs.  This is, of  course, a challenging research problem, but beyond the scope of our present work.  However, our present work should be useful for understanding the small but finite amplitude electrostatic disturbances  in many space (viz. noctilucent clouds \cite{Markus99-NLCs}, Earth's mesosphere \cite{Havnes96,Gelinas98,Mendis04,Mahmoudian13}, cometary tails \cite{Horanyi96}, Jupiter's surroundings \cite{Tsintikidis96}, Jupiter's magnetosphere \cite{Horanyi93}, etc.) and laboratory\cite{Samarian01-lab,Khrapak01-lab,Fortov03-ed-lab} dusty plasma environments,  where the electron, ion pcd plasma species coexist. 
\section*{Data Availability Statement}
Data sharing is not applicable to this article as no new data were created or analyzed in this study.

\end{document}